\documentclass[prd,nofootinbib,preprint,superscriptaddress,twocolumn,10pt]{revtex4}
\usepackage{amsmath,amssymb}
\usepackage{soul}
\usepackage{epsfig}
\usepackage{graphicx}
\usepackage[usenames,dvipsnames]{color}
\usepackage{slashed}
\usepackage[colorlinks,citecolor=blue]{hyperref}
\usepackage{color}
\def\beq{\begin{equation}}
\def\eeq{\end{equation}}
\def\bey{\begin{eqnarray}}
\def\eey{\end{eqnarray}}

\def\lsim{\mathrel{\raise.3ex\hbox{$<$\kern-.75em\lower1ex\hbox{$\sim$}}}}
\def\gsim{\mathrel{\raise.3ex\hbox{$>$\kern-.75em\lower1ex\hbox{$\sim$}}}}

\newcommand{\mx}{m_\chi}
\newcommand{\ax}{\alpha_\chi}
\newcommand{\mphi}{m_\phi}
\def\beq{\begin{equation}\begin{aligned}}
\def\eeq{\end{aligned}\end{equation}}

\begin{document}
\title{Self-interacting Dark Matter in Non-Standard Cosmology}

%\author{abc}
\author{Manoranjan Dutta}
\email{dutta.manoranjan@nlc.ac.in}
\affiliation{Department of Physics, North Lakhimpur College (Autonomous), Khelmati, North Lakhimpur, Lakhimpur, Assam 787031, India}

%%%%%%%%%%%%%%%%%%%%%%%%%%%%%%%%

%\input mylatex

%\doublespacing
%\textwidth 15.55cm \textheight 22.5cm
%\hoffset -2cm
%\voffset -1cm

%\def\jw#1{{\color{blue}{#1}}}
%\def\jwn#1{{\color{red}{\bf[#1]}}}

%================================================
%==================================================

%\begin{document}
%\maketitle
%\newpage
%\title{Asymmetric Self-interacting Dark Matter via Dirac Leptogenesis}

%\author{abc}
%\author[a]{Manoranjan Dutta}
%\emailAdd{ph18resch11007@iith.ac.in}

%\author[b]{Nimmala Narendra}
%\emailAdd{nnarendra@prl.res.in}

%\author[a]{Narendra Sahu}
%\emailAdd{nsahu@phy.iith.ac.in}

%\author[c,d,c]{Sujay Shil}
%\emailAdd{sujay.s@iitgn.ac.in}

%\affiliation[a]{Department of Physics, Indian Institute of Technology Hyderabad}

%\affiliation[b]{Theoretical Physics Division, Physical Research Laboratory, Ahmedabad - 380009, India}

%\affiliation[c]{Department of Physics, Indian Institute of Technology Hyderabad}

%\affiliation[c]{IIT Gandhinagar, Palaj Campus, Gujarat 382355, India}
%\affiliation[d]{Institute of Physics, Sachivalaya Marg, Bhubaneswar, Pin-751005, Odisha}
%\affiliation[e]{Homi Bhabha National Institute, BARC Training School Complex, Anushakti Nagar, Mumbai 400094, India}

%\keywords{}
%%%%%%%%%%%%%%%%%%%%

%\flushbottom

%%%%%%%%%%%%%%%%%%%%%%%%%%%%%%%%%%%%%%%%%%%%%%%%%%%%%%%%%%%%%%%%%%%%%
  
%%%%%%%%%%%%%%%%%%%%%%%%%%%%%%%%%%%%%%%%%%%%%%%%%%%%%%%%%%%%%%%%%%%%%%%  
\begin{abstract}
Discrepancies of the $\Lambda {\rm CDM}$ model with small-scale cosmological observations and stringent constraints from direct search experiments cast doubts over typical weak scale cold dark matter candidates {\it e.g.} WIMPs. Self-interacting dark matter (SIDM) is a very promising alternative to WIMP, which not only alleviates the small-scale anomalies of the $\Lambda {\rm CDM}$ model, but also matches with the highly accurate large-scale $\Lambda$CDM predictions. The small-scale anomalies can be resolved with a self-scattering cross-section $\sigma/m_{DM} \sim 1 cm^2/gm$. Such large cross-sections can be realised in models of DM with a light MeV scale mediator and a sizeable coupling. We assume the DM to be Dirac fermion and the mediator to be either a light scalar or vector boson. In a standard cosmological history, one major issue with such models is to realise the correct relic density of dark matter via thermal freeze-out as the DM annihilates efficiently to the light mediators and ends up with an under-abundant relic. However, we show that, if the expansion rate of the universe is not radiation-dominated (RD) during the epoch of SIDM freeze-out, its relic abundance is enhanced significantly. We assume a non-standard expansion history of the universe by introducing a non-radiation like component in the early universe. In such a scenario, DM freezes out at an earlier epoch resulting in enhanced DM abundance, which can be matched with the correct relic density with suitable model parameters. The light mediator can also couple to an SM mediator and pave a way to detect SIDM at terrestrial laboratories. The mixing between the mediators can be constrained by data from direct search experiments. We find out the viable parameter space for a generic SIDM model taking into account the relevant phenomenological and experimental constraints.

\end{abstract}

\maketitle
\newpage

%%%%%%%%%%%%%%%%%%%%%%%%%%%%%%%%%%%%%%%%%%%%%%%%%%%%%%%%%%%%%%%%%%%%%%%
%\preprint{ HRI-RECAPP-2020-007}
%\keywords{Dark matter, asymmetric dark matter, self-interacting dark matter, leptogenesis, neutrino mass.}

%\begin{document}

%\maketitle
%\flushbottom

\section{Introduction} \label{Intro}
%%%%%%%%%%%%%%%%%%%%%%%%%%%%%%%%%%%%%%%%%%%%5
%%%%%%%%%%%%%%%%%%%%%%%%%%%%%%%%%%%%%%%%%%%%%%%%%%%%%%%%%%%%%%%%%

Irrefutable evidence from galaxy rotation curves, gravitational lensing, cosmic microwave background {\it etc.} suggests the existence of a non-luminous and non-baryonic form of matter in the universe, known as dark matter (DM)~\cite{Zwicky:1933gu, Rubin:1970zza, Clowe:2006eq}. Satellite-borne experiments like Planck and WMAP, which measured the anisotropies in the cosmic microwave background (CMB), predict that DM makes up about $26.8\%$ of the present energy density of the universe, which also amounts to nearly 85\% of the total matter of the present universe. DM abundance is conventionally expressed as~\cite{Aghanim:2018eyx}: $\Omega_{\text{DM}} h^2 = 0.120\pm 0.001$ at 68\% CL, where $\Omega_{\rm DM}$ is the DM density parameter and $h = \text{Hubble Parameter}/(100 \;\text{km} ~\text{s}^{-1} \text{Mpc}^{-1})$ is the reduced Hubble parameter. As no standard model (SM) particle qualifies to be a DM particle, one needs to go beyond the standard model to explain DM. Over the past several decades, the `weakly interacting massive particles (WIMPs)' have become very popular as potential DM candidates. WIMPs are stable particles with mass and couplings in the weak scale which were assumed to be in thermal equilibrium with the SM bath in the early universe and subsequently frozen out to yield a relic abundance that matches with Planck or WMAP measurements. This interesting coincidence is often referred to as the `WIMP miracle'. However, the null detection of DM in the terrestrial DM search experiments casts doubt over the minimal WIMP scenario. After probing DM-nucleon cross-section down to $10^{-47} {\rm cm^2}$~\cite{Aprile:2018dbl}, direct search experiments rule out typical WIMPs up to a few TeVs. 
%Very soon, direct search experiments will reach the sensitivity where the coherent elastic neutrino-nucleus scattering (CE$\nu$NS) will give an irreducible background to the DM signal-also known as the `{\it neutrino floor}'. 
Non-observation of missing energy signature at the Large Hadron Collider (LHC) also put constraints on minimal WIMP-like DM models.

\begin{figure}[htb!]
\centering
%\includegraphics [scale=0.3] {dm-int.png}
%\hfill
\includegraphics [scale=0.12] {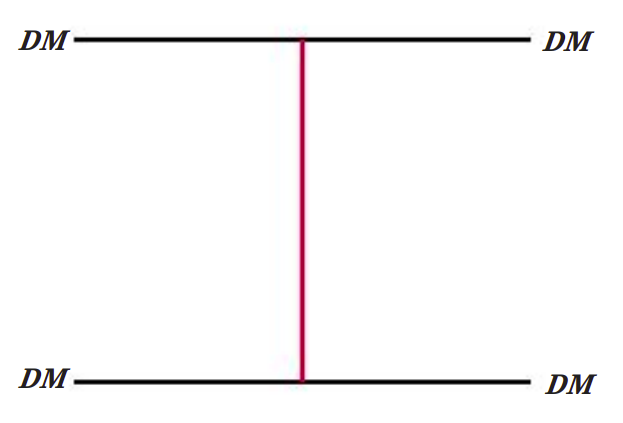}
%\hfill
\includegraphics [scale=0.12] {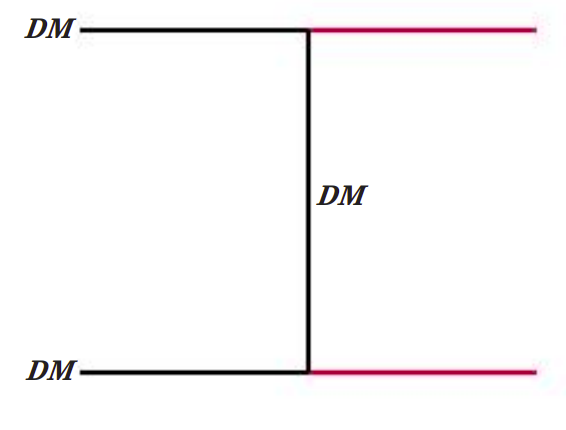}
%\hfill
\includegraphics [scale=0.14] {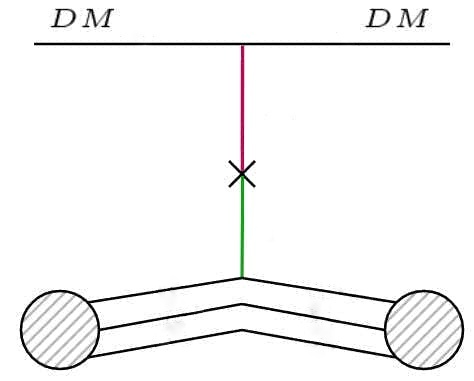}
\caption{%Left: $2 \to 2$ DM-SM effective interactions in WIMP framework that yields relic density, direct search, indirect search and collider cross-sections. Right: 
\footnotesize{Left: Feynman graph for DM self-interaction, Middle: Feynman graph for dominant DM annihilation, Right: Feynman graph for direct detection of SIDM. The black and red lines depict SIDM and light BSM mediators respectively and the green line represents an SM mediator. The specific properties of these particles depend on the choice of model.}} 
\label{feyn}
\end{figure}

From the cosmological point of view, the Standard Model of Cosmology or the ${\rm \Lambda CDM}$ model assumes DM to be a cold and collisionless fluid. WIMPs owing to their non-relativistic nature and very small cross-section, fit well to be cold and collisionless. The ${\rm \Lambda CDM}$ model has been enormously successful in explaining the large-scale structures of the universe. However, at small scales, more prominently at the scale of dwarf galaxies, quite a few discrepancies of ${\rm \Lambda CDM}$ predictions have emerged {\it e.g.} the core-cusp problem, the missing satellite problem and the too-big-to-fail problem {\it etc}~\cite{Tulin:2017ara, Bullock:2017xww}. To alleviate these anomalies, Spergel and Steinhardt \cite{Spergel:1999mh} proposed a new variant of DM  as an alternative to conventional collision-less CDM of $\Lambda{\rm CDM}$ model that goes by the name of self-interacting dark matter (SIDM); see~\cite{Carlson:1992fn, deLaix:1995vi} for earlier studies. To alleviate the anomalies, the self-scattering cross-section required is of the order $\sigma/m \sim 1 \; {\rm cm}^2/{\rm g} \approx 2 \times 10^{-24} \; {\rm cm}^2/{\rm GeV}$ \cite{Buckley:2009in, Feng:2009hw, Feng:2009mn,Loeb:2010gj, Zavala:2012us, Vogelsberger:2012ku}, which is many orders of magnitude larger than typical WIMP annihilation cross-section ($10^{-36}{\rm cm}^2/{\rm GeV}$). Astrophysical observations also require the self-interaction to be stronger at smaller DM velocities and thus have a larger impact on small scale structures while being in agreement with CDM predictions at larger scales (typically at cluster scale and beyond) with larger DM velocities \cite{Buckley:2009in, Feng:2009hw, Feng:2009mn, Loeb:2010gj, Bringmann:2016din, Kaplinghat:2015aga, Aarssen:2012fx, Tulin:2013teo}. Such a velocity-dependent self-interaction can be realised in SIDM models with a light BSM mediator. Optimistically one can assume that there exists some coupling of this light BSM mediator with some SM mediator which can be probed at direct search experiments~\cite{Kaplinghat:2013yxa, DelNobile:2015uua}. Several model building efforts have been made to realise such scenarios, see~\cite{Kouvaris:2014uoa, Bernal:2015ova, Kainulainen:2015sva, Hambye:2019tjt, Cirelli:2016rnw, Kahlhoefer:2017umn,Dutta:2021wbn,Borah:2021yek,Borah:2021pet} and references therein. However, due to sizeable $\mathcal{O}(1)$ coupling between SIDM and the light mediator, the SIDM annihilates very efficiently into light mediators via t-channel annihilation and its thermal relic becomes under-abundant. We show in Fig.~\ref{feyn}, the Feynman graphs for DM self-interaction (left), dominant annihilation channel (middle) and direct search.

The important point to be noted in this regard is that the vast majority of the literature assumes that DM decouples during the epoch of radiation domination, although there is no indispensable reason for this to be true. If DM froze out when the universe was dominated by a non-radiation like component, the Hubble parameter of the universe would change which will change the freeze-out dynamics~\cite{Evans:2019jcs,Hamdan:2017psw,Allahverdi:2019jsc,Allahverdi:2018aux, Drees:2017iod, Davoudiasl:2015vba,Delos:2019dyh, Randall:2015xza, Co:2015pka, Dror:2016rxc, DEramo:2017gpl, DEramo:2017ecx, Garcia:2018wtq, Bernal:2018kcw, Drees:2018dsj, Betancur:2018xtj, Maldonado:2019qmp, Poulin:2019omz, Arias:2019uol, Bramante:2017obj, Bernal:2018ins, Hardy:2018bph, Kaneta:2019zgw, Bernal:2019mhf, Berlin:2016vnh, Kamionkowski:1990ni,Chanda:2019xyl,Konar:2020vuu,Ghosh:2021wrk}. Some earlier studies in this direction include DM decoupling in inflationary reheating \cite{McDonald:1989jd,Chung:1998rq,Giudice:2000ex,Gelmini:2006pw} or kination domination \cite{Spokoiny:1993kt,Ferreira:1997hj,Salati:2002md,Pallis:2005hm} {\it etc}. Here in this work, we explore matter-dominated freeze-out of SIDM and explore whether the correct thermal relic of SIDM can be realised in the light of matter-dominated freeze-out.

This paper is structured as follows: In section \ref{sec2} we outline a generic minimal SIDM model with a light vector/scalar mediators and in section~\ref{sec3} we discuss the freeze-out abundance of SIDM in a matter-dominated universe and compare it to the radiation dominated case. In section \ref{sec3}, we explore the direct detection constraints on the model parameters followed by constraints from big-bang nucleosynthesis (BBN) in section \ref{sec4} and indirect detection in section \ref{sec5}. Finally, we summarise and conclude in~\ref{sec6}.  
%%%%%%%%%%%%%%%%%%%%%%%%%%%%%%%%%%%%%%%%%%%%%%%%%%%%%%%%%%%%%%%%%%%%%%%%%%%%%%%%%%%%%%%%%%%%%%%%%%%
\section{Dark Matter Self-interaction} \label{sec2}
%%%%%%%%%%%%%%%%%%%%%%%%%%%%%%%%%%%%%%%%%%%%%%%%%%
SIDM with light mediators can be realised easily in BSM models extended by scalar or vector boson mediators~\cite{Buckley:2009in, Feng:2009hw, Feng:2009mn,Ibe:2009mk,Loeb:2010gj, Bringmann:2016din, Kaplinghat:2015aga, Aarssen:2012fx,Tulin:2012wi, Tulin:2013teo,Schutz:2014nka}. Spontaneously broken $U(1)$ models are well suited in this direction, where the stability of DM is assured due to charge conservation.  With a sizeable coupling and light mediator, the desired self-scattering cross-section can be obtained.
The interaction Lagrangian for such models assuming the DM to be a Dirac fermion $\chi$ is given by:
\beq
{\cal L_{\rm int}}=\bigg\{
\begin{array}{c l}
g_D\bar{\chi}\gamma^\mu\chi\phi_\mu & \text{(vector mediator)}\\
g_D\bar{\chi}\chi\phi & \text{(scalar mediator)} \, 
\end{array}
\eeq
where $\phi_\mu$ ($\phi$) is the vector (scalar) mediator, and $g_D$ is the dark coupling constant. Self-scattering, in the non-relativistic limit, is well-described by the Yukawa-type potential,
\beq
V(r)=\pm\frac{\alpha_D}{r}e^{-\mphi r} 
\label{eq:yukawa}
\eeq 
where $\alpha_D = \frac{g^2_D}{4\pi}$.
In case of a vector mediator, $\chi \overline{\chi}$ scattering leads to a attractive ($-$) potential, while $\chi \chi$ or $\bar\chi \bar\chi$ scattering leads to a repulsive ($+$) potential. In the case of a scalar mediator, the potential is always attractive.
To capture the relevant physics of forward scattering divergence for the self-interaction, one can define the transfer cross-section $\sigma_T$ as~\cite{Feng:2009hw,Tulin:2013teo,Tulin:2017ara}
\begin{equation}
	\sigma_T = \int d\Omega (1-\cos\theta) \frac{d\sigma}{d\Omega}
\end{equation}
 Depending on the masses of DM ($m_\chi$) and the mediator ($m_{\phi}$), as well as the relative velocity of the colliding particle ($v$) and the coupling ($\alpha_D$), we can identify three distinct regimes. The perturbative Born regime ($\alpha m_\chi/m_\phi \ll 1,  m_\chi v/m_{\phi} \geq 1$) is where the perturbative calculation holds good. Outside the Born regime, we have the classical regime ($\alpha_D m_\chi/ m_\phi \geq 1, m_\chi v/m_{\phi} \geq 1$) and the resonant regime ($\alpha_D m_\chi/ m_\phi \geq 1, m_\chi v/m_{\phi} \leq 1$) where non-perturbative and quantum-mechanical effects become important. The self-interaction cross-sections in these regimes are listed in Appendix~\ref{appendix2}. In Fig.~\ref{sidm1}, we show the self-interaction allowed parameter space in $m_\chi$ - $m_{\phi}$ plane obtained by constraining $\sigma/m_\chi$ in the correct ballpark from astrophysical data across different scales.  We constrain $\sigma/m_\chi$ in the range $0.1-10~{\rm cm}^2/{\rm g}$ for galaxies ($v\sim 100~ \rm km/s$) and dwarf galaxies ($v\sim 10~ \rm km/s$) shown by the shades of cyan and green coloured region as indicated in the figure inset. The light magenta coloured region depicts the parameter space allowed for clusters ($\sigma/m_\chi \sim 0.1-1~{\rm cm}^2/{\rm g}$). The masses of DM and the mediator for which all three regions overlap will alleviate the small-scale anomalies across all scales.	
	\begin{figure}[h!]
	\centering
		\includegraphics[width=7.5cm,height=7cm]{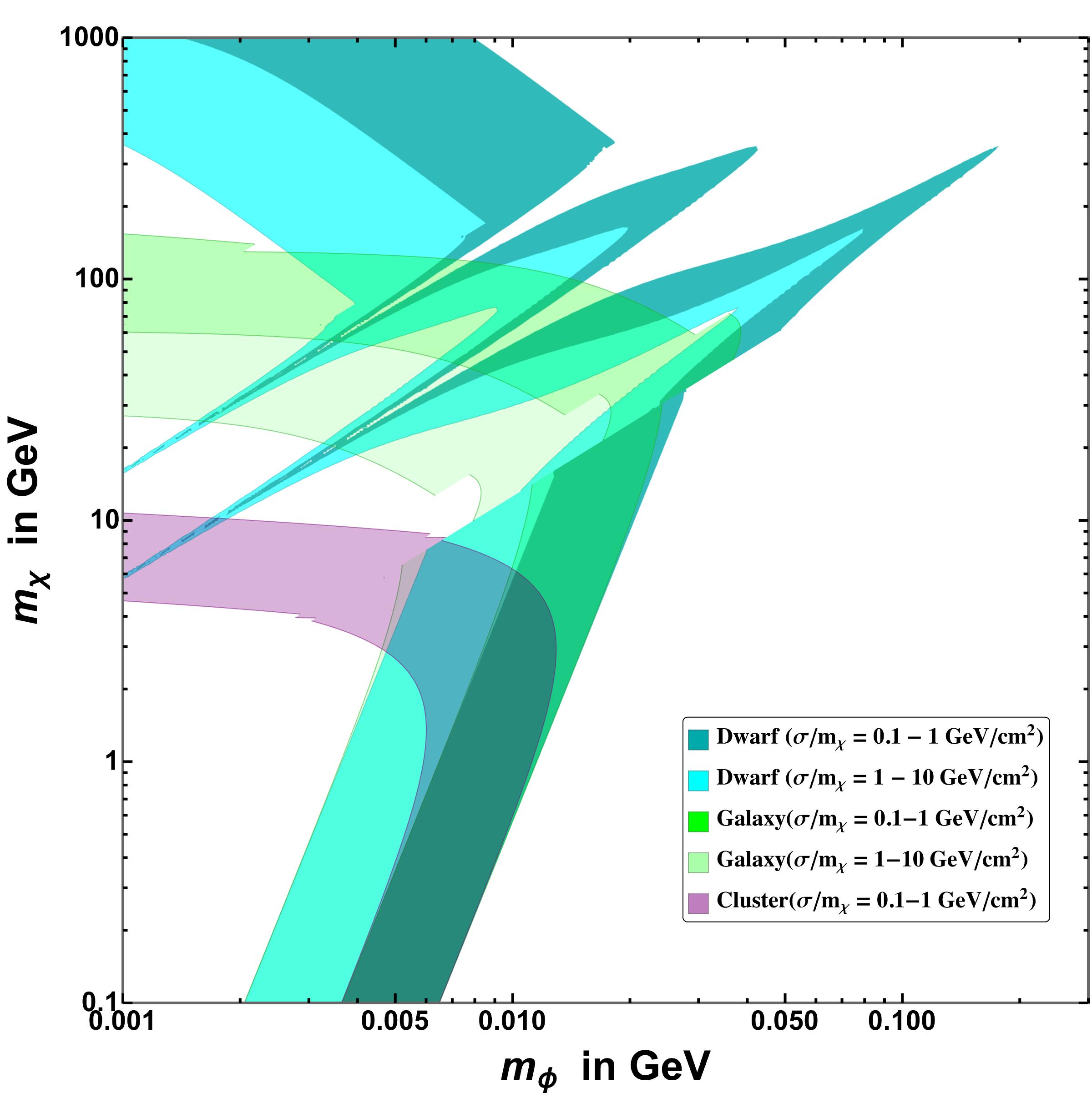}
		%\hfill
		%\vspace*{1cm}
		%\includegraphics[width=7.5cm,height=9cm]{sidm_astro}
		\caption{Self-interaction cross-section in the range $0.1-1 \; {\rm cm}^2/{\rm g}$ for clusters ($v\sim1000 \; {\rm km/s}$), $0.1-10 \; {\rm cm}^2/{\rm g}$ for galaxies ($v \sim 100 $ km/s.) and $0.1-10 \; {\rm cm}^2/{\rm g}$ for dwarfs ($v \sim 10 $ km/s).}
		\label{sidm1}
	\end{figure}	
	The top (bottom) corner corresponds to the Classical (Born) region. In these two regimes, the dependence of the cross-section on velocity is trivial. The region sandwiched between these two regions is called the resonant region, where quantum mechanical resonances and anti-resonances appear due to (quasi)bound state formation in the attractive potential. The resonances are most prominent at the dwarf scale as DM velocity dispersion is lower at the dwarf scale and gradually becomes less prominent towards galaxy and cluster scales as DM velocity increases. For a coupling $g_D$, the condition $m_\chi v/m_\phi < 1$ dictates the onset of non-perturbative quantum mechanical effects, which is easily satisfied at smaller velocities. We have considered in Fig.~\ref{sidm1}, $g_D = 0.1$ which gives the self-scattering cross-section at the correct ballpark. We have checked that for $g_D < 0.02$, the obtained cross-sections are below the ballpark of $\sigma/m_\chi \sim 0.1 cm^2/g$, insufficient to alleviate the small scale $\Lambda{\rm CDM}$ anomalies.

	%The resonant spikes are not distinct in these figures as we have varied the Yukawa coupling in a range 0.1-1. Nevertheless, prominent resonant spikes can be seen in Fig.~\ref{sidmdd} in section~\ref{sec5}, where we show the same parameter space for a fixed Yukawa coupling $y'_1=0.35$, while confronting the SIDM parameter space to direct search. We can see from the figures that a wide range of DM mass can give rise to sufficient self-interaction. However, the mass of the mediator is constrained roughly within two orders of magnitudes excepting for the resonance case. 
	%We will confront these regions of parameter space to direct detection bounds in section~\ref{sec5}.
	
\begin{figure}
	\includegraphics[width=7.5cm,height=9cm]{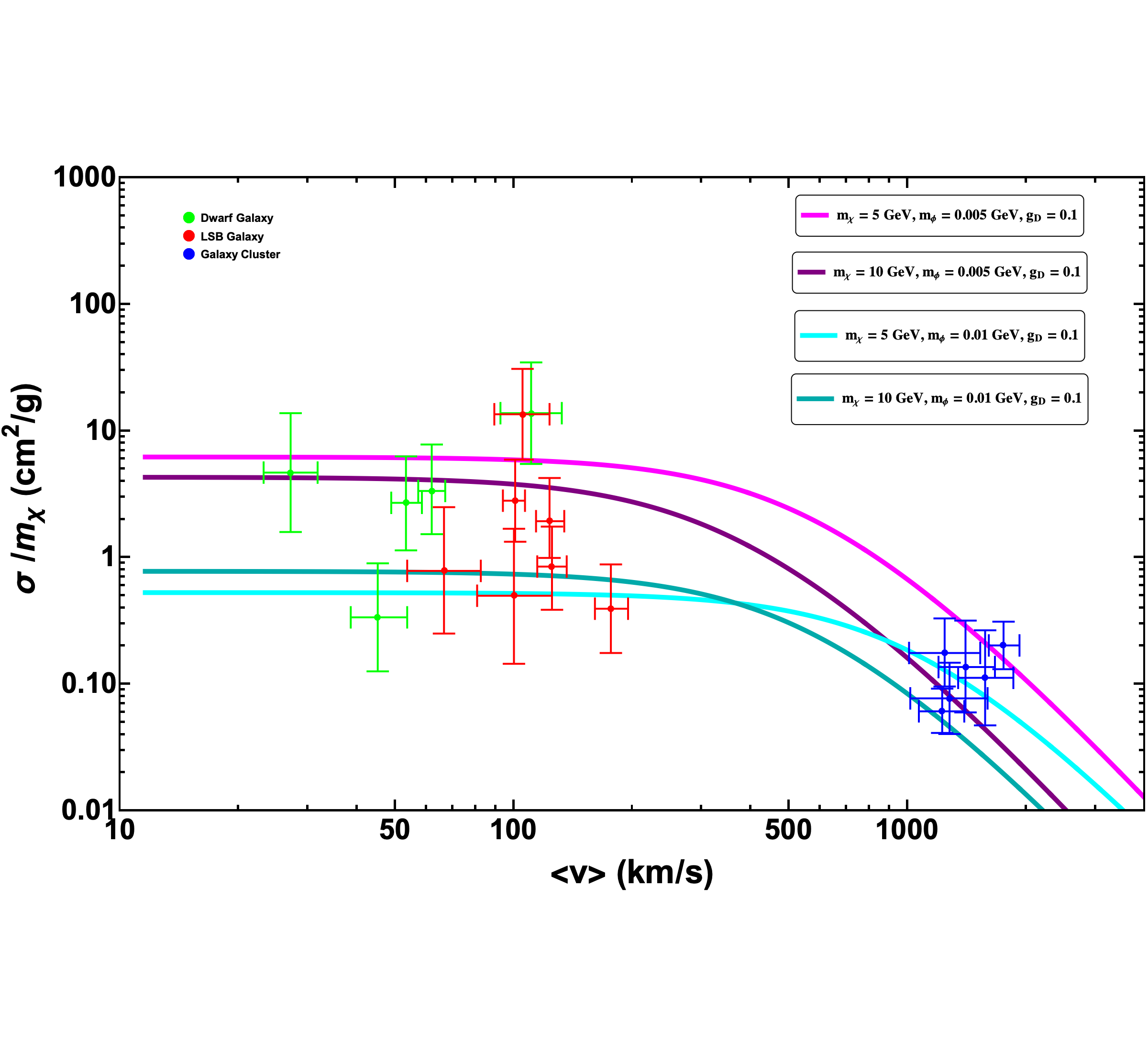}
\caption{The self-interaction cross section per unit mass of DM as a function of average collision velocity.}
\label{sidm2}
\end{figure}

	We also plot the self-scattering cross-section per unit DM mass as a function of average collision velocity, which fits the available data from dwarfs (red), low surface brightness (LSB) galaxies (blue), and clusters (green)~\cite{Kaplinghat:2015aga, Kamada:2020buc} as shown in Fig \ref{sidm2}. Different coloured curves represent different combinations of $m_\chi$ and $m_\phi$ keeping the coupling fixed at $g_D = 0.1$. It is clear from Fig.~\ref{sidm2} that the model can appreciably explain the astrophysical observation of velocity-dependent DM self-interaction.	

%The annihilation cross-section mainly depends on the free parameters: $\Phi-H$ nixing, {\it i.e.} $\sin \beta$, the mass of $h_{2}${\it i.e.} $M_{h_{2}}$ and the $\chi$ coupling with $h_{2}$, {\it i.e.} $\lambda_{DM}$. These free parameters are highly constrained by invisible Higgs decay~\cite{Khachatryan:2016whc}, relic density of DM reported by WMAP~\cite{Hinshaw:2012aka} and Planck~\cite{Akrami:2018vks}, SI direct detection cross-sections at XENON100~\cite{Aprile:2012nq}, LUX~\cite{Akerib:2016vxi}, XENON1T~\cite{Aprile:2015uzo}\, and the Higgs signal strength measured at LHC~\cite{cms_report_2018, Khachatryan:2016vau}.

\section{Relic Density of SIDM in Non Standard Cosmology}\label{sec2}

\begin{figure}[h!]
		\centering
		\includegraphics[width=8cm,height=8cm]{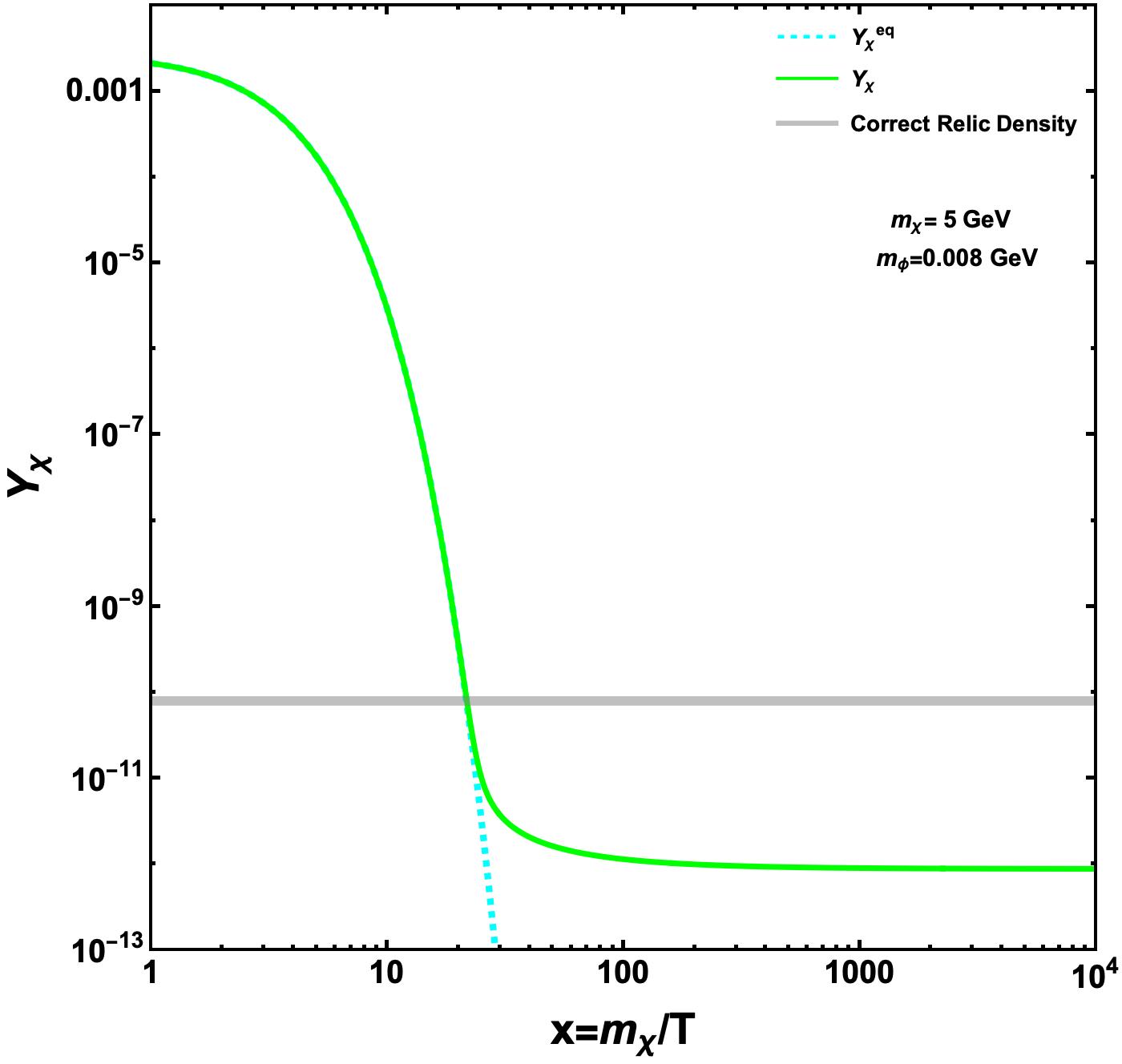}
		\caption{Thermal freeze out of SIDM in non-standard cosmology}	
			\label{relic_stand}
	\end{figure}

In the standard cosmology, the dark matter decoupling happens in the radiation domination. In this scenario, the Hubble parameter ($H$) is given by
	\begin{equation}\label{hubble_s}
		H^2=\frac{8\pi}{3M_{\rm pl}^2}\left(\rho_R+\rho_{\chi}\right)
	\end{equation}
	where  $\rho_R$ and $\rho_{\chi}$ are, respectively, the energy densities of the SM radiation bath and DM. Assuming the DM to be in the thermal universe in the early universe, it freezes out from the cosmic soup dominantly via the t-channel Feynman diagram shown in the middle panel of Fig.~\ref{feyn}. While the `{\it DM-DM-Mediator}' should be sizeable to realise sufficient self-interaction, the same sizeable coupling appears in its dominant annihilation channel, which leads to a cross-section which is much larger than the one governing typical WIMP annihilation. As a consequence, the thermal freeze-out abundance is well below the correct ballpark. The thermally averaged cross-section for this annihilation diagram can be approximated as,
	\beq
\langle\sigma v\rangle = \Bigg\{
\begin{array}{c l}
\frac{\pi\ax^2}{\mx^2}\sqrt{1-\frac{\mphi^2}{\mx^2}} & \text{(vector mediator)}\\
\frac{3}{4}\frac{\pi\ax^2}{\mx^2}v^2\sqrt{1-\frac{\mphi^2}{\mx^2}} & \text{(scalar mediator)}
\end{array}
\label{cross}
\eeq
		where $\alpha_D=g^2_D/(4\pi)$ and $m_\chi (m_\phi)$ is the mass of DM (mediator). As evident from Eqns.~\ref{cross}, the annihilation to vector (scalar) mediators is a s-wave (p-wave) process. While we take into account the Sommerfeld enhancement while discussing the indirect detection in Sec.~\ref{inddet}, it is found to be negligibly small at the epoch of DM decoupling. Also due to very high velocity, both cross-sections can be assumed to be almost equal at the epoch of DM decoupling, which is not true for the present epoch relevant for indirect detection. To study, the evolution of the DM number density, $n_\chi$ we use the standard form of the Boltzmann equation,
	\begin{equation}
	\dot n_\chi +3Hn_\chi=-\langle\sigma v\rangle[n_{DM}^2-(n_{DM}^{\rm eq})^2]
	\label{BE}
	\end{equation}
For convenience, we define the dimensionless parameter $x=m_{\chi}/T$ and the co-moving DM number density as $Y_{\chi}=n_{\chi}/s(T))$ where, $s(T)==\frac{2\pi^2}{45}g_{* S}T^3$. In terms of $Y_{\chi}$ and $x$, we write down the Boltzmann equations as,
		\begin{equation}
		\label{b_sn}
\frac{dY_{\chi}}{dx}= -\frac{s(m_{\chi})}{x^2  H(m_{\chi})} \langle\sigma v \rangle \left(Y^2_{\chi} -\left(Y^{eq}_{\chi}\right)^2\right)
\end{equation}
In Fig.~\ref{relic_stand}, we show the freeze-out abundance of standard freeze-out using Eq.~\ref{b_sn} for a DM of mass 5 GeV. The mass of the mediator does not affect the magnitude of the cross-section appreciably as long as $m_{\phi}\ll m_{\chi}$, yet for concreteness, we consider $m_{\phi}=0.008$ GeV. As evident from the figure, the relic is under-abundant by orders of magnitude. 

\begin{figure}[h!]
		\centering
		\includegraphics[width=8cm,height=8cm]{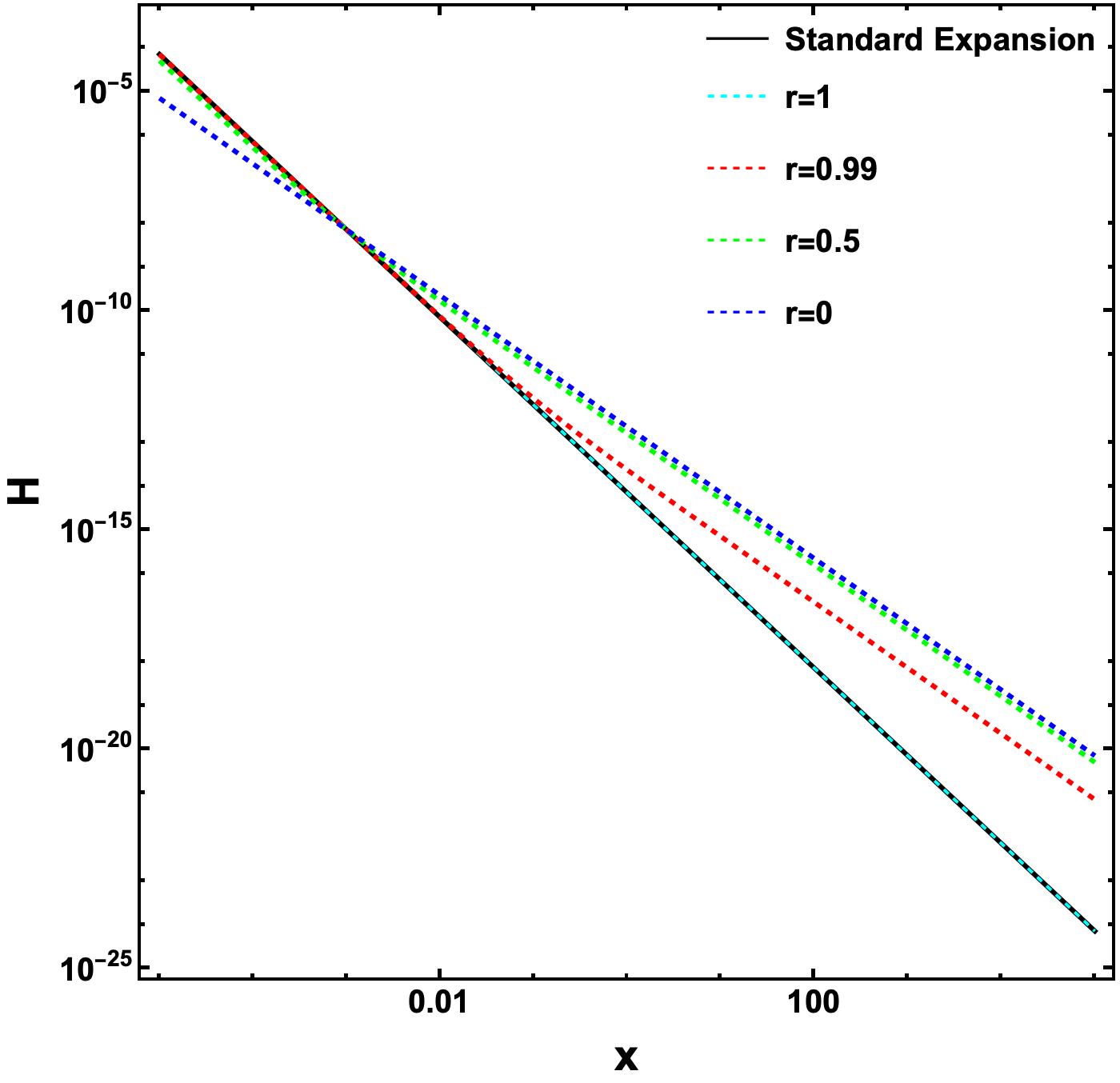}
		\caption{Comparison of Modified Hubble parameter with the standard Hubble parameter. }	
			\label{fig_hubble}
	\end{figure}

As the correct thermal relic of SIDM can not be realised in the standard freeze-out scenario, we invoke a non-standard freeze-out scenario considering that, in the early universe, there exists a heavy scalar $\eta$ with energy density $\rho_\eta$ besides the components considered in the standard cosmology. 
%In the simplest case, $\phi$ is some heavy species that decouples from the thermal bath at $T_\star\sim m_\phi$ and becomes matter-like. 
%\footnote{More generally if $\phi$ is never in thermal contact with the thermal bath then $T_\star$ has considerable freedom since it denotes the temperature of the Standard Model bath at which $\phi$ becomes matter-like.} 
In the presence of this extra component, The Hubble parameter ($H$) is given by,
	\begin{equation}\label{hubble}
		H^2=\frac{8\pi}{3M_{\rm pl}^2}\left(\rho_R+\rho_{\chi}+\rho_{\eta}\right)
	\end{equation}
We define a critical temperature $T_\star$, such that for $T>T_\star$, it evolves like radiation ($\phi \sim a^{-4}$) while for $T<T_\star$, it evolves like matter ($\phi \sim a^{-3}$). For $T_\star>T\gg m_Z,m_\chi$, Eqn.~\ref{hubble} can be rewritten as,
	\begin{equation}\label{FEhS}
\frac{H^2}{H_\star^2}=\frac{g_* r}{g_*+g_{\chi}}\left(\frac{a_\star}{a}\right)^{4}+(1-r)\left(\frac{a_\star}{a}\right)^{3}+\frac{g_{\chi}r}{g_*+g_{\chi}}\left(\frac{a_\star}{a}\right)^{4}
	\end{equation}
where, $g_{*}$, $g_\chi$ and $g_\eta$ are the effective number of relativistic degrees of freedom (DoF) of the SM, internal DoF of DM and internal DoF of $\eta$ respectively. The quantity $r$ represents the fraction of the energy in radiation at temperature $T_{\star}$ and hence $(1-r)$ represents the fraction of energy in the $\eta$ component at $T=T_{\star}$. The quantity $r$ can be defined as,
\beq
	r\equiv\frac{\rho_R+\rho_{\chi}}{\rho_R+\rho_{\chi}+\rho_\eta}\Bigg|_{T=T_\star}~.
	\label{2.4}
\eeq
In terms of $T_\star$, $H_\star$ can be re-expressed as,
	\begin{equation}\label{eq:FEf}
		H_\star^2= \frac{8\pi^3}{90M_{{\rm pl}}^2}\big(g_{*}(T_\star)+g_{\chi}+g_\eta\big)T_\star^{4}
	\end{equation}
	This form of Friedmann equation signifies a non-standard cosmological history where the universe was initially radiation dominated ($H \propto T^2$) for $r \sim 1$, but becomes matter dominated ($H\propto T^{\frac{3}{2}}$) at $T_\star$ for $r\ll 1$. 
	%Here $r$ is the ratio of energy densities at the initial time and is not a function of time. A good benchmark for $r$ (which we use at various places below) is $r\simeq0.99$ this corresponds to the case that $\phi$ has $\mathcal{O}(1)$ internal DoF, the DM and Standard Model combined have $\mathcal{O}(100)$ DoF and that all states shared a common temperature in the past which has not significantly deviated by  $T=T_{\star}$.
Assuming that the entropy is conserved in the SM bath and that $g_\star$ remains constant until DM decoupling, the scale factor is related to the temperature as,
\beq
\label{aeq}
\left(\frac{a_\star}{a}\right)\simeq \left(\frac{T}{T_\star}\right) 
\eeq
%
%Note, however, that $\phi$ decays eventually lead to entropy non-conservation in the Standard Model bath thus invalidating the relationship of eq.~(\ref{aeq}). Once entropy is no longer conserved in the bath, one instead has a scenario reminiscent to  \cite{McDonald:1989jd,Chung:1998rq,Giudice:2000ex,Gelmini:2006pw} in which freeze-out occurs during a period of entropy injection to the thermal bath, similar to the period of inflationary reheating. We will return to this point shortly and discuss the transition. 	
Neglecting small contribution from DM and making use of dimensionless variables $x\equiv m_{\chi}/T$ and $x_{\star}\equiv x(T_{\star})$, the Friedmann equation can be written as
	\begin{equation}\label{eq:FEx}
		H \equiv H_\star \left(\frac{x_\star}{x}\right)^\frac{3}{2} \left(r \frac{x_\star}{x} + 1 - r\right)^\frac{1}{2}
	\end{equation}
	In Fig.~\ref{fig_hubble}, we compare the modified Hubble parameter with the standard Hubble parameter for different values of $r$. We see that for $r=1$, the modified Hubble parameter matches with the Standard Hubble parameter and it deviates from the standard parameter as $r$ decreases from 1. 

%As $\rho_\eta$ starts evolving as matter at $T_\star$, its fractional contribution to the total energy density grows until it eventually dominates the energy density of the universe.
% We will discuss the requirement on the $\phi$ lifetime to realise this scenario shortly.  
%
%We define the freeze-out temperature, $T_{f}\equiv \frac{m_{\chi}}{x_{f}}$ as the temperature at which $\Gamma_{\rm ann}(x_{f}) = H(x_{f})$ where, $\Gamma_{\rm ann$ is

\begin{figure}[h!]
		\centering
		\includegraphics[width=8cm,height=8cm]{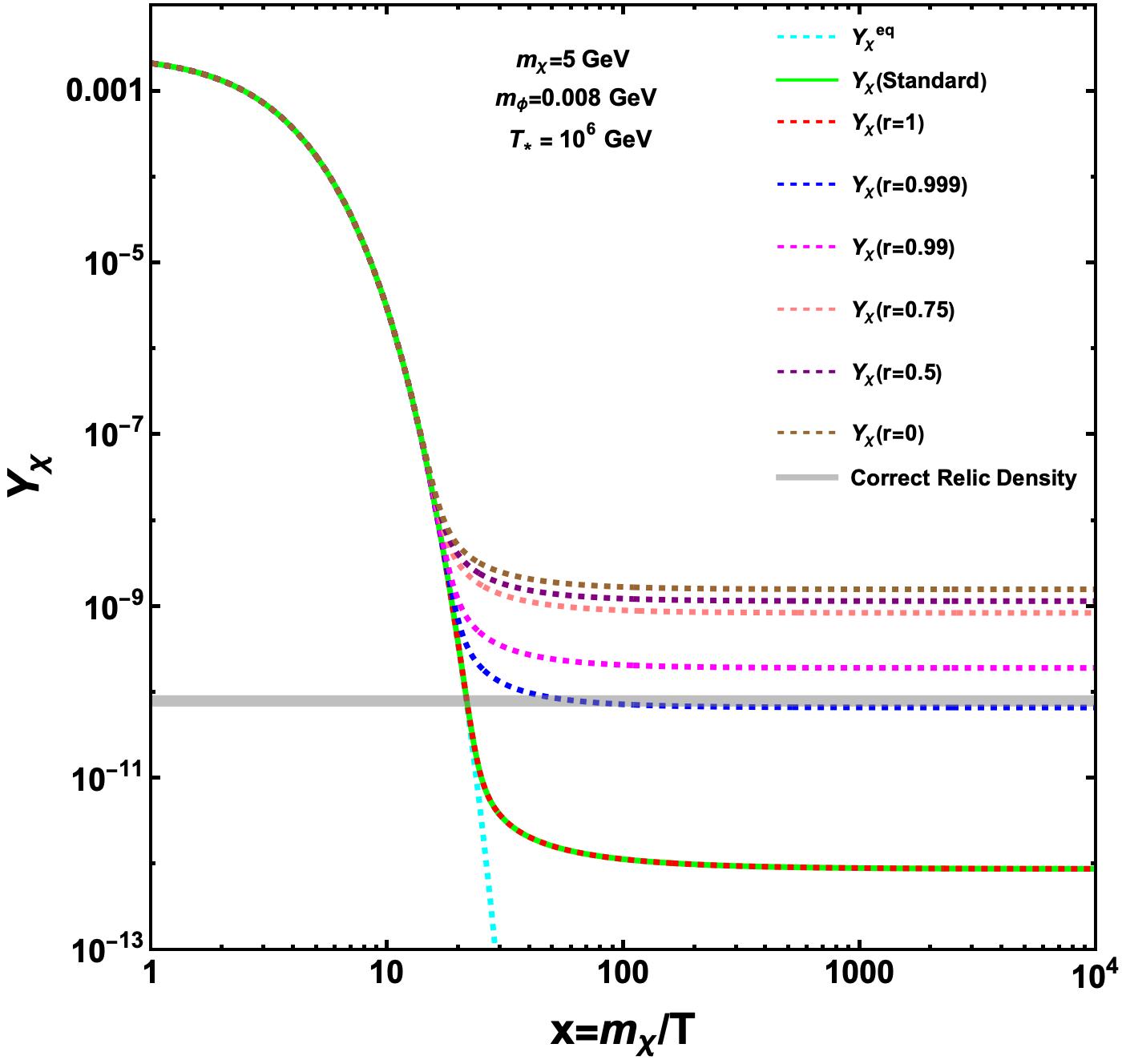}
		\caption{Thermal freeze out of SIDM in non-standard cosmology}	
			\label{sidm_ns}
	\end{figure}

	%In this scenario, if the DM decouples from the bath at en epoch where $r\ll 1$, so that $H\propto T^{\frac{3}{2}}$ unlike the universe dominated case ($H\propto T^{2}$), the freeze-out abundance of DM would be different from the radiation dominated freeze-out.
%%%%%%%%%%%%%%%%%%%%%%%%%%%%%%%%%%%%%%%%%55
In the light of these modifications, the Boltzmann Eqn. can be written as,  
\beq
	\frac{dY}{dx}=-\frac{s(m_{\chi})  \langle\sigma v\rangle }{x^2 H(m_{\chi}) \left(\frac{x}{x_\star} \right)^{\frac{1}{2}} \left(1-r+r\frac{x_\star}{x}\right)^\frac{1}{2}}  \left(Y^2_{\chi}-\big(Y^{\rm eq}_{\chi}\big)^2\right)
	\eeq
%where $\langle\sigma v\rangle$ is the thermally averaged cross-section of DM annihilation into light mediators, given by,
%\beq
 %\left(\sigma_{\rm an} v\right)^{\rm tree}_V=\frac{\pi\ax^2}{\mx^2}\sqrt{1-\frac{\mphi^2}{\mx^2}},~\left(\sigma_{\rm an} v\right)^{\rm tree}_S=\frac{3}{4}\frac{\pi\ax^2}{\mx^2}v^2\sqrt{1-\frac{\mphi^2}{\mx^2}}
% \eeq
\label{eq:ann}
As desired, in the limit $r=1$, we get back to the Boltzmann equation in standard cosmology (Eqn. \ref{BE}). The relic abundance for a DM of mass 5 GeV using the modified Boltzmann equation is shown in Fig.~\ref{sidm_ns}, with $T_\star = 10^{6}$ GeV and $m_{\phi}=0.008$ GeV. As we can see from Fig.~\ref{sidm_ns}, the relic abundance significantly increases as $r$ decreases from 1 gradually. The $r=1$ case exactly matches the abundance in the standard case, as expected. The maximum relic abundance is obtained for $r=0$ depicting maximum matter domination. 

Note that, the universe has to return to radiation domination before the onset of BBN in order to keep BBN predictions intact, and this can be ensured by the decay of extra component $\eta$ into SM particles, which injects extra entropy into the radiation bath. This also means that, after the DM freezes out, its relic abundance is diluted by a factor $\zeta$, which parametrises the change in the entropy of the radiation bath due to $\eta$ decay. Assuming the sudden decay approximation, the magnitude of the entropy injection can be quantified as,
 \beq
\label{eq:ZetaTRH}
 \zeta \equiv \frac{s_{\rm before}}{s_{\rm after}} \simeq \left(\frac{T_{\Gamma_\eta}}{T_{\rm RH}}\right)^3~.
\eeq
where $T_{\rm RH}$ is the Reheating temperature of the universe and $\Gamma_\eta$ is the decay width of $\eta$. In terms of the Planck mass and the decay width of $\eta$, we can define,
\beq
T_{\rm RH} \equiv \sqrt{\Gamma_{\eta}M_{Pl}} \gtrsim10 {\rm MeV}
\label{trh}
\eeq
Here, we neglect any changes in degrees of freedom. $T_{\Gamma_\eta}$ is obtained by evolving $T_\star$ using $(a_\star/a_{\Gamma_\eta})$, where $a_{\Gamma_\eta}$ is defined by the condition $H(a_{\Gamma_\eta})\simeq\Gamma_\eta$. 
Assuming that $\eta$ decays after it dominates the energy density, using eqn.~\ref{eq:FEx}, 
%The ratio of scale factors can be expressed in terms of the initial condition $a_\star$ and the point $a_{\Gamma}$ where $H=\Gamma_{\phi}$,
\beq\label{Th1c}
  \frac{a_{\star}}{a_{\Gamma}} \approx \frac{1}{(1-r)^{1/3}}\left(\frac{\Gamma_{\eta}}{H_{\star}}\right)^{2/3} 
  \eeq
Then using eq.~\ref{aeq} and eq.~\ref{trh}, we obtain,
\beq
T_{\Gamma_{\phi}}& \simeq \left(\frac{45}{4\pi^{3}g_*(1-r)}\frac{T_{{\rm RH}}^{4}}{T_{\star}}\right)^{1/3}
 \eeq
 Substituting $\zeta$ as a function of $T_{\rm RH}$, as in eq.~\eqref{eq:ZetaTRH}, we get,
 \beq\label{eq:TRHOmega}
 \zeta \simeq \frac{45}{4\pi^{3}}   (1-r)  \frac{g_\star T_{\star}}{T_{\rm RH}}
 \eeq
Due to this change in entropy, there is a dilution in the DM relic density. Therefore, the final relic is slightly lower than the freeze-out relic. They are related as,
	\begin{equation}
	Y_{\rm relic}=\frac{n_\chi}{s_{\rm after}}=\zeta\frac{ n_\chi}{s_{\rm before}}=\zeta Y_{\rm FO} .
	\end{equation}
It follows that, the final DM relic abundance is,
\beq
\Omega_{\chi}^{\rm relic}=\zeta\Omega_{\chi}^{\rm FO} = \zeta\times\frac{s_0m_\chi Y_{\rm FO}}{\rho_c}
\eeq 
where $\rho_c$ and $s_0$ are the critical density and the present entropy density of the universe respectively.

\begin{figure}[h!]
		\centering
		\includegraphics[width=8cm,height=8cm]{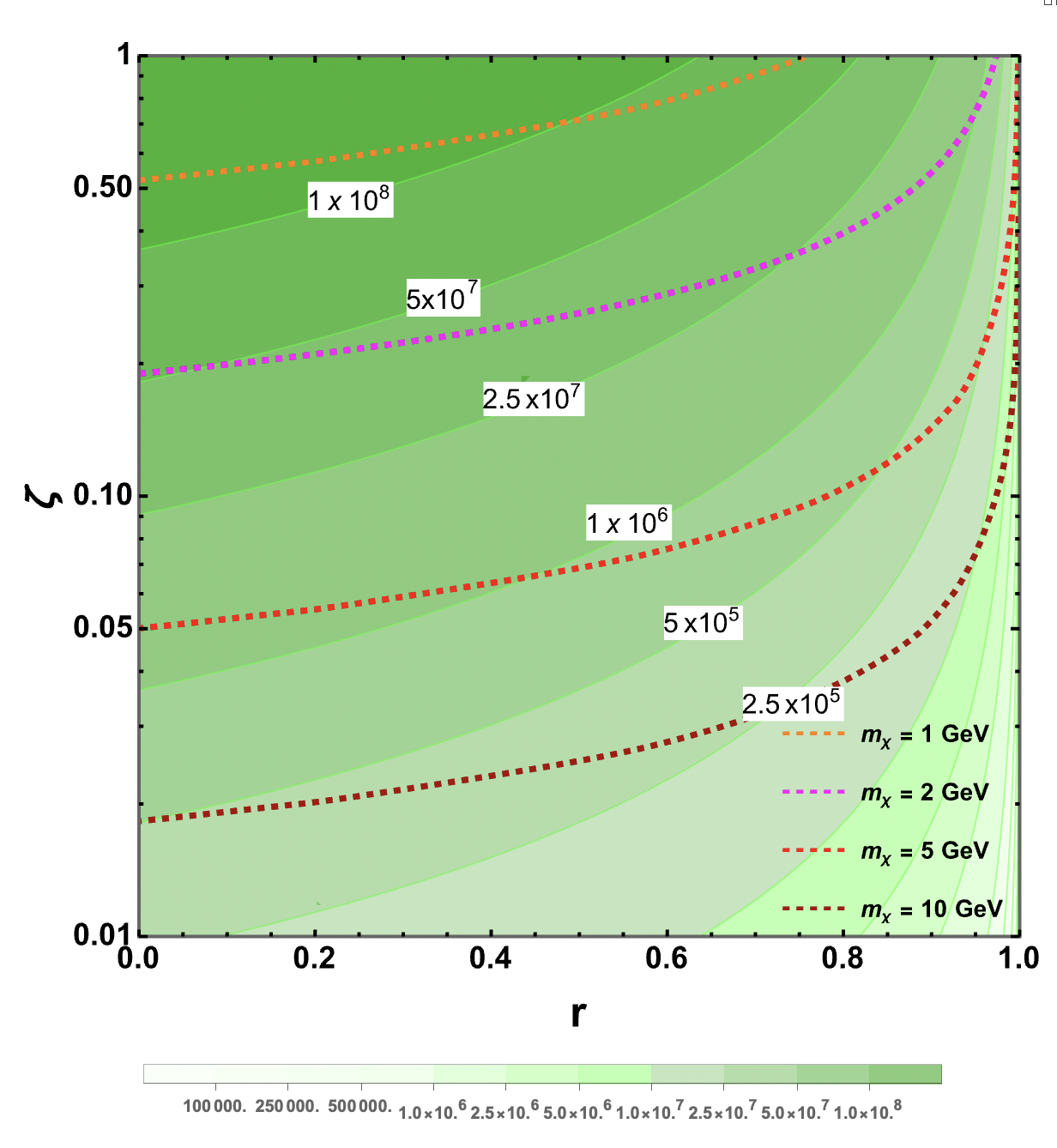}
		\caption{The $\zeta$ vs $r$ plot for different DM masses, along with contours of $T_{\rm RH}/{\rm GeV}$.}	
			\label{plot}
	\end{figure}

In Fig.~\ref{plot}, we show the required dilution factor $\zeta$ corresponding to different $r$ values for different DM masses (as indicated in colour codes in the figure inset) in order to match the observed DM relic density. We have also shown, in the same plane, the contours of $T_{\rm RH}/{\rm GeV}$ using eq.~\ref{eq:TRHOmega} depicted in green colour with the value indicated inside white boxes. The colour-coded scale for $T_{\rm RH}/{\rm GeV}$ is also shown below the plot. Note that $\zeta > 1$ means the DM relic obtained from freeze-out is already under-abundant, and that is why the region with $\zeta > 1$ is not included in the plot. Therefore, it is evident that the correct relic abundance for SIDM masses suitable for alleviating small-scale anomalies can be obtained via its freeze-out in this modified cosmological scenario.

\section{Constraints from Direct Search} \label{sec3}
SIDM can be detected in terrestrial laboratories through the mixing between the light BSM mediator and some SM bosons. If the light BSM mediator is scalar, then there arises the possibility of scalar mixing with SM Higgs ($\phi-h$), while if the light mediator is a vector boson, it can mix with the SM Z boson {\it i.e.,} kinetic mixing is between $U(1)_D$ and $U(1)_Y$. In the latter case, the BSM vector boson can mix with photon as well, however, the mixing with Z is stronger due to larger coupling.

For gauge sector kinetic mixing, the spin independent direct search cross section for a given nucleus N with proton number Z and mass number A is given by,
\begin{equation}
\sigma_{\rm SI}^{\phi-Z}=\frac{g^2 g^2_D \epsilon^2}{\pi}\frac{ \mu^2_{\chi N}}{M^4_{Z'}} \frac{\big( Z f_p + (A-Z) f_n \big)^2}{A^2}
\end{equation}
where, $\mu_{\chi N} = \frac{m_\chi m_N}{(m_\chi+m_N)}$ is the reduced mass of the DM-nucleus system, $m_{n}$ being the nucleon (proton or neutron) mass, $\epsilon$ is the kinetic mixing, A and Z are respectively the mass and atomic number of the target nucleus. 

In case of scalar mixing ($\theta_{\phi h}$), the scattering cross-section of DM per nucleon can be expressed as,
	\begin{equation}
		\sigma_{SI}^{\phi-h} = \frac{\mu_{\chi N}^{2}}{4\pi A^{2}} \left[ Z f_{p} + (A-Z) f_{n} \right]^{2}
		\label{DD_cs}
	\end{equation}
where, $f_{p}$ and $f_{n}$ are the interaction strengths of proton and neutron with DM respectively, given as,
	\begin{equation}
		f_{p,n}=\sum\limits_{q=u,d,s} f_{T_{q}}^{p,n} \alpha_{q}\frac{m_{p,n}}{m_{q}} + \frac{2}{27} f_{TG}^{p,n}\sum\limits_{q=c,t,b}\alpha_{q} 
		\frac{m_{p,n}}{m_{q}}
		\label{fpn}
	\end{equation}
	where 
	\begin{equation}
		\alpha_{q} =  \lambda_D \theta_{\phi h}\left( \frac{m_{q}}{v}\right) \left[\frac{1}{m^2_\phi}-\frac{1}{m^{2}_h}\right] 
		\label{DD4}
	\end{equation}

		\begin{figure}[h!]
		\centering
		\includegraphics[width=8cm,height=8cm]{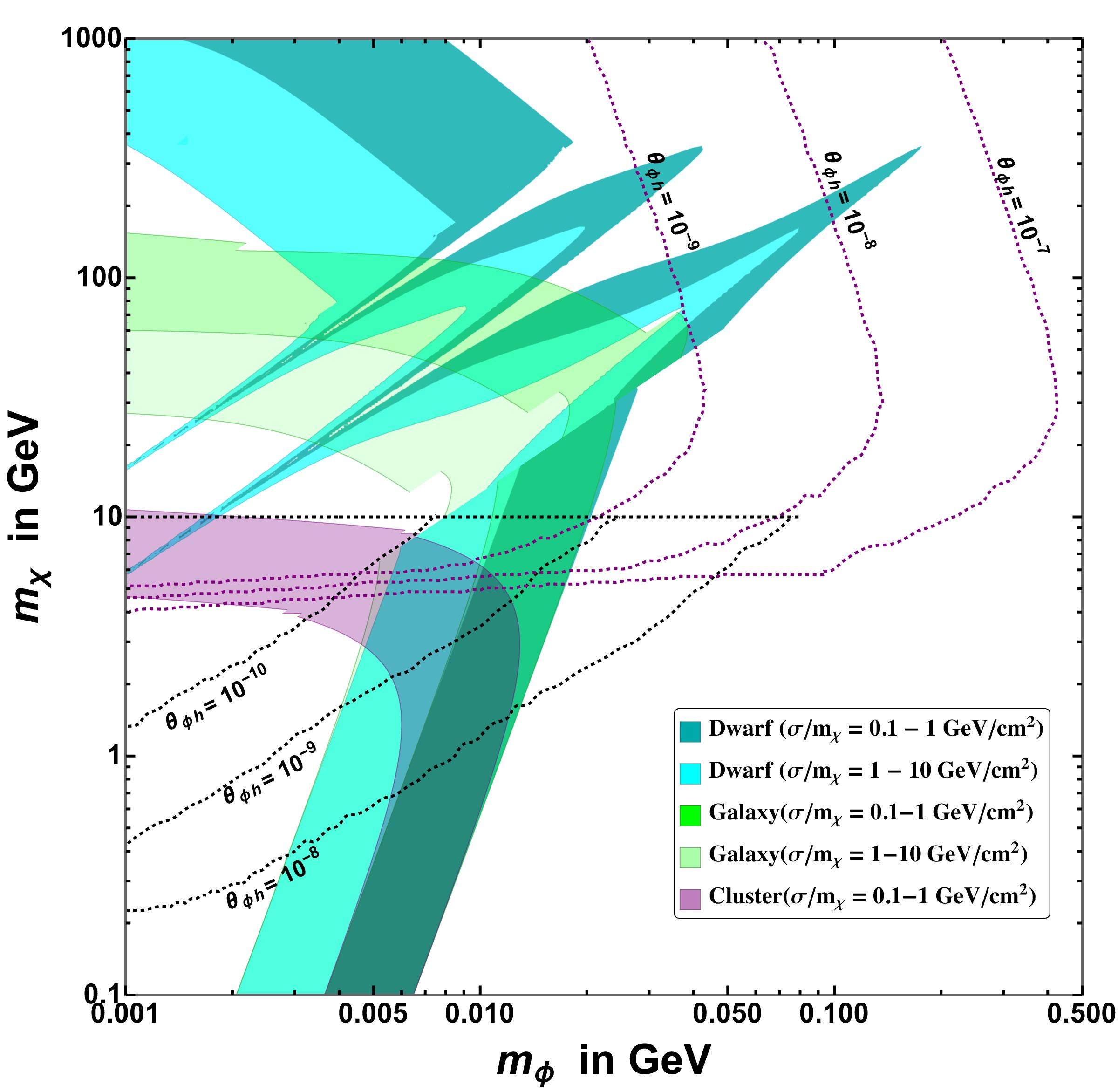}
		\caption{ Self-interaction allowed parameter space constrained by DM direct search for the vector mediated case.}	
			\label{sidmdd_scalar}
	\end{figure}
	
		\begin{figure}[h!]
		\centering
		\includegraphics[width=8cm,height=8cm]{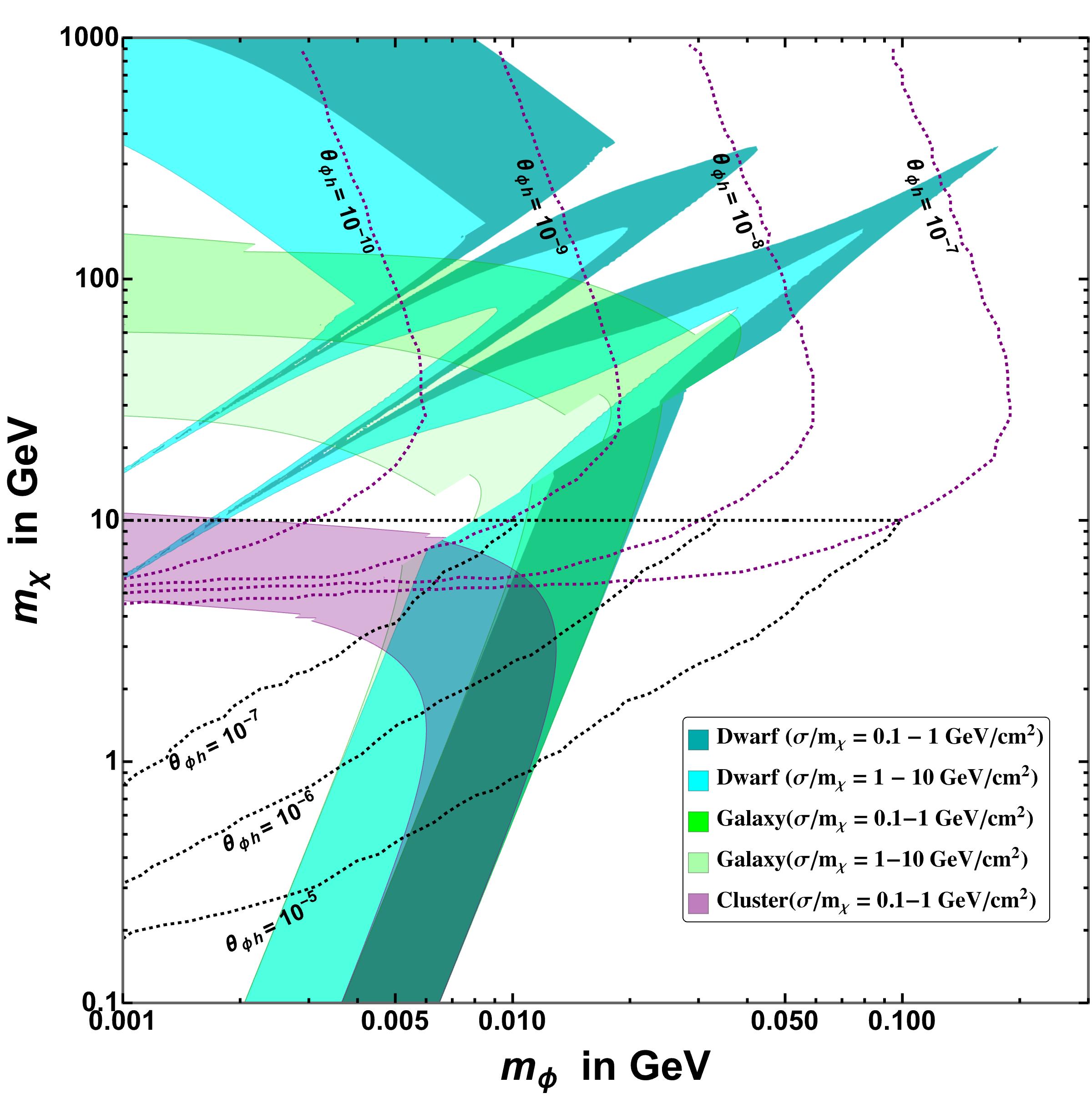}
		\caption{ Self-interaction allowed parameter space constrained by DM direct search for the scalar mediated case.}	
			\label{sidmdd_boson}
	\end{figure}

As mentioned earlier, DM direct search experiments like CRESST-III~\cite{Abdelhameed:2019hmk} and XENON1T \cite{Aprile:2018dbl} can constrain the model parameters. While XENON1T provides the most stringent bound for DM of mass above 10 GeV, CRESST constraints the below 10 GeV mass range. In Fig.~\ref{sidmdd_boson} and Fig.~\ref{sidmdd_scalar}, the most stringent constraints from CRESST-III~\cite{Abdelhameed:2019hmk}, XENON1T \cite{Aprile:2018dbl} experiments on $m_\chi-m_{\phi}$ plane are shown against the parameter space favoured from required DM self-interactions by assuming $g_D=0.1$. The dotted purple-coloured contours denote exclusion limits from the XENON1T experiment for specific kinetic mixing parameters such that the region towards the left of the contour is excluded for that particular kinetic mixing. Similarly, the dotted black coloured contours show the CRESST-III bound on low mass DM for different kinetic mixing parameters ruling out the parameter space towards the left of each contour.

	%Among the direct search experiments, CRESST-III~\cite{Abdelhameed:2019hmk} provides the most severe constraint on DM mass below 10 GeV, while XENON1T \cite{Aprile:2018dbl} provides the stringent constraints for DM mass above 10 GeV. In Fig.~\ref{sidmdd}, these constraints are shown on the $m_{\chi}-m_\phi$ plane against the self-interaction favoured parameter space. The blue (purple) coloured contours denote exclusion limits from XENON1T (CRESST-III) experiment for specific $\phi-h$ mixing parameter $\theta_{\phi h}$. The region to the left of each contour is excluded for that particular $\theta_{\phi h}$. It is seen from Fig.~\ref{sidmdd} that direct search experiments severely constrain the self-interaction favoured parameter space. In particular, for $m_{\chi}=7.76$ GeV and $m_\phi=0.01$ GeV, $\theta_{\phi h} > 10^{-9}$ has already been ruled out. The red star mark depicts the benchmark point for fully asymmetric DM.
\section{Constraints from Big Bang Nucleosynthesis}	\label{sec4}
\begin{figure}[h!]
		\centering
		\includegraphics[width=8cm,height=8cm]{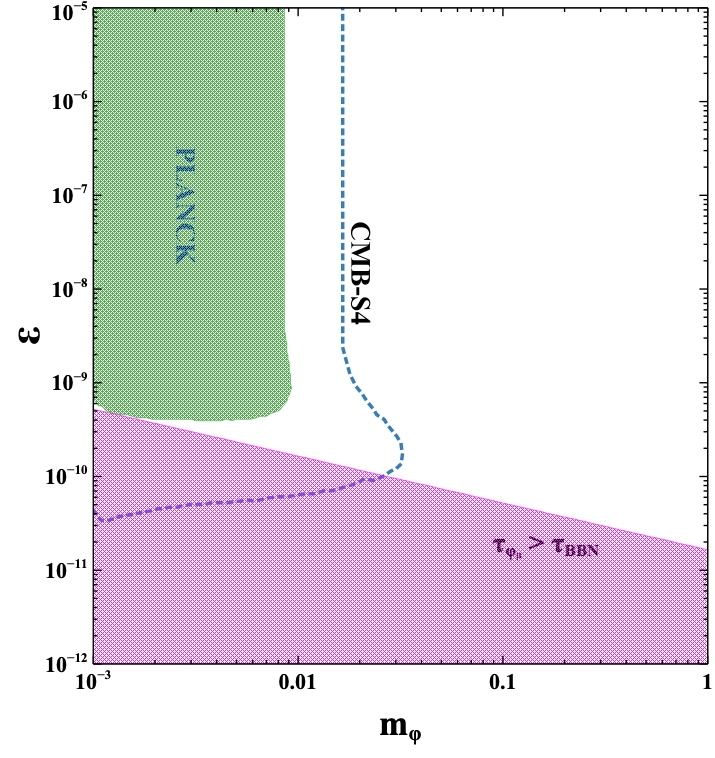}
		\caption{ $\epsilon$ versus light vector mass ($m_{\phi}$) confronted with BBN bounds.}	
			\label{bbn_boson}
	\end{figure}

\begin{figure}[h!]
		\centering
		\includegraphics[width=8cm,height=8cm]{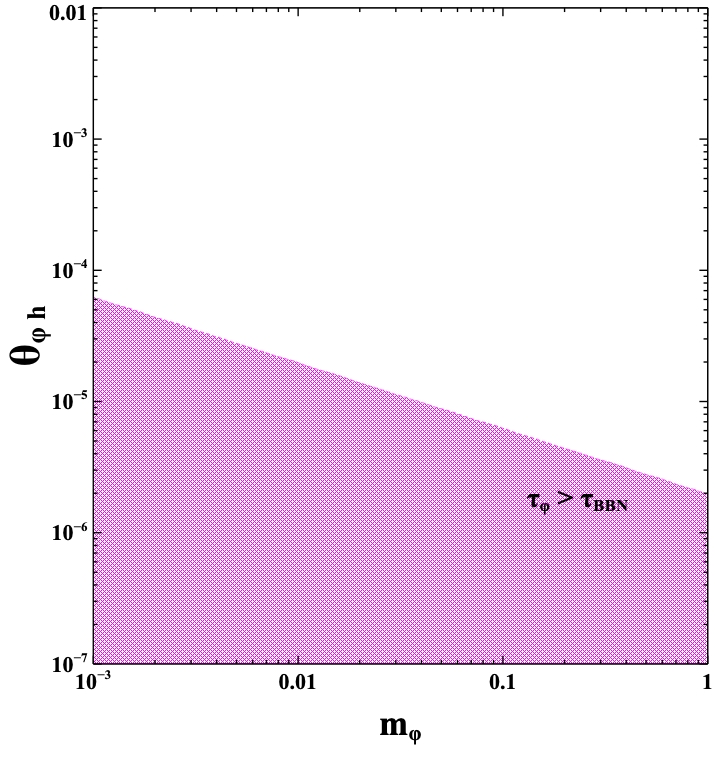}
		\caption{$\theta_{\phi h}$ versus light scalar mass ($m_{\phi}$) confronted with BBN bounds.}	
			\label{bbn_scalar}
	\end{figure}

	Late DM annihilations produce mediator pairs copiously, which can potentially spoil the BBN prediction about light element abundance in the present universe. Therefore, we apply a conservative bound on the lifetime of the mediators to be less than the typical BBN epoch so that BBN predictions remain intact. In Fig.~\ref{bbn_boson}, kinetic mixing parameter $\epsilon$ is shown against $m_{\phi_\mu}$. Very small kinetic mixing (the cyan-coloured region) is disfavoured from this conservative lifetime bound. We show the analogous bound for the scalar mediator case in Fig.~\ref{bbn_scalar}. Comparing Fig.~\ref{bbn_scalar} and Fig.~\ref{sidmdd_scalar}, we see that the scalar mediator case is almost ruled out except for a tiny parameter space below DM mass of 1 GeV (see Fig.~\ref{sidmdd_scalar}). The vector mediator case has ample parameter space safe from BBN bounds. The green region in Fig.~\ref{bbn_boson} is excluded from cosmological constraints on effective relativistic degrees of freedom \cite{Aghanim:2018eyx}, while the dotted blue line shows the projected sensitivity of CMB-S4 experiment\cite{CMB-S4:2016ple,Ibe:2019gpv}.
	%The decay width of these processes is summarized in Appendix~\ref{appen1}.

\section{Constraints from Indirect Search}\label{sec5}
\label{inddet}
As discussed in Sec.~\ref{sec2}, the annihilation channels of DM to vector mediators is an S-wave process, where the cross-section is independent of velocity ($\sigma v \sim v^0$) while that to scalar mediators is a p-wave process, hence velocity suppressed ($\sigma v \sim v^2$).
Since the mediators, whether vector or scalar, are very light, the annihilation cross-section must be multiplied by the Sommerfeld enhancement factor~\cite{https://doi.org/10.1002/andp.19163561802}, which reflects the modification of the initial-state wave function due to multiple mediator exchange: $(\sigma v)_\text{enh} = S \times \sigma v$. The Sommerfeld enhancement factors for $s$-wave and $p$-wave annihilations are given respectively as~\cite{Cassel:2009wt,Iengo:2009ni,Slatyer:2009vg},
\begin{equation}
	\begin{aligned}
		S_s&=\frac{\pi}{a}\frac{\sinh(2\pi a c)}{\cosh(2\pi ac)-\cos(2\pi\sqrt{c-(ac)^2})}\\
		S_p&=\frac{(c-1)^2+4(ac)^2}{1+4(ac)^2} S_s
	\end{aligned}
\end{equation}
where $a=2\pi v/g^2$ and $c=3 g^2 m_\chi/2\pi^3 m_\phi$

%for an $s$-wave annihilation process~\cite{Cassel:2009wt,Iengo:2009ni,Slatyer:2009vg} 
%\begin{equation}
% S_s = \frac{\pi}{a} \frac{\sinh (2 \pi \, a \, c)}{\cosh (2 \pi \, a \, c) - \cos (2 \pi \sqrt{c - a^2 c^2})} \; ,
%\end{equation}
%where $a = v/(2 \alpha_\psi)$ and  $c = 6 \, \alpha_\psi \, m_\psi / (\pi^2 m_\phi)$ with $\alpha_\psi = y_\psi^2 \, \cos^2 \delta_\psi / (4\pi)$
%\footnote{Note that we only use the scalar part of the coupling to calculate the Sommerfeld enhancement. We will return to this issue in more detail in the context of DM self-interactions in section~\ref{sec:constraints} and in particular in appendix~\ref{app:pseudoSIDM}.} The corresponding expression for a $p$-wave process is
%\begin{equation}
% S_p = \frac{(c-1)^2 + 4 \, a^2 c^2}{1 + 4 \, a^2 c^2} \times S_s \; . 
%\end{equation}
For $v \gtrsim \frac{g^2}{4\pi}$, we get $S_{s,p} \approx 1$ (and hence not significant at the epoch of DM freeze-out), whereas for smaller velocities $S$ goes as $1/v$ in the $s$-wave case and $1/v^3$ in the $p$-wave case, so that effectively the annihilation cross sections for both cases goes as $1/v$. The enhancement saturates for $v \lesssim m_\phi / (2 m_{\chi})$, hence the ratio of the two masses determines the maximum possible enhancement. Since all the annihilations to SM final states are further suppressed by the kinetic or scalar mixing, effectively $\langle \sigma v \rangle_{{\rm DM~DM} \to {\rm SM~SM}} \sim \epsilon^2/v$ in case of the vector-mediated case and  $\langle \sigma v \rangle_{{\rm DM~DM} \to {\rm SM~SM}} \sim \theta^2_{\theta-h}/v$ in case of the scalar mediated case. Thus, all fluxes of gamma rays, cosmic rays and neutrinos are well below the present and future reach of indirect detection probes~\cite{Arina:2018zcq}, even with the maximum possible Sommerfeld enhancement. For example, for $m_{\chi} = 10~ {\rm GeV}, m_\phi = 10~ {\rm MeV}, g = 0.1$, the Sommerfeld enhancement factor is of $\mathcal{O}(1)$ and hence the effective annihilation cross-section of DM to SM final states is far below the current limits from Fermi-LAT~\cite{Fermi-LAT:2015att,Fermi-LAT:2013sme}, MAGIC~\cite{MAGIC:2016xys}, HESS~\cite{HESS:2018cbt}, AMS-02~\cite{PhysRevLett.110.141102}, constraints from CMB by Planck~\cite{Aghanim:2018eyx} and $ \gamma $-rays by INTEGRAL~\cite{Knodlseder:2007kh}.

%\section{Final Parameter space}	
%%%%%%%%%%%%%%%%%%%%%%%%%%%%%%%%%%%%%%%%%%%%%%%%%%%%%%%%%%%%%%%%%%%%%%%%%%%%%%%%%%%%%%%%%%%%%%%%%%%
\section{Conclusion}\label{sec6}

In this study, we investigate the viability of achieving the correct relic abundance for self-interacting dark matter (SIDM) in the context of non-standard cosmological scenarios. Our analysis focuses on a generic SIDM model, employing a Dirac fermion as the SIDM candidate and considering either a scalar or boson mediator to enable the self-interaction.

In the case of standard freeze-out during the radiation-dominated epoch, the resulting DM relic abundance is found to be under-abundant. However, the freeze-out dynamics changes substantially in the presence of additional components in the early universe. We propose a scenario wherein the expansion history of the universe is altered due to the presence of an extra component, resulting in a considerable enhancement of the dark matter relic through the early decoupling of SIDM from the thermal bath. Furthermore, we account for entropy changes arising from the decay of this extra field into Standard Model (SM) particles prior to the onset of Big Bang Nucleosynthesis (BBN).

We constrain the parameter space favoured for self-interactions, considering experimental data from both direct and indirect searches, as well as BBN constraints. The permissible parameter space for self-interactions is sandwiched between the upper bounds from direct searches and the lower bounds from BBN. Intriguingly, our analysis indicates that the scalar-mediated SIDM model is largely disfavoured by these constraints collectively. Conversely, the vector-mediated SIDM model displays a substantial parameter space that remains viable for exploration in future experiments.
Given the commencement of numerous light DM search experiments in the coming decade, coinciding with the natural alignment of our SIDM scenario within this category, we anticipate a pivotal period for testing the SIDM paradigm.
% The next decade holds significant promise for advancing our understanding of the fundamental nature of dark matter through dedicated investigations into self-interacting dark matter scenarios.

%%%%%%%%%%%%%%%%%%%%%%%%%%%%%%%%%%%%%%%%%%%%%%%%%%%%%%%%%%%%%%%%%%%%%%%%%%%%%%%%%%%%%%%%%%%%%%%%%%%

\appendix

%\section*{Appendix}

\section{DM Self-interaction Cross-sections at Low Energy}
	\label{appendix2}
	In the Born Limit ($\lambda^2_D m_{\chi}/(4\pi m_\phi) << 1$),
	\begin{equation}
		\sigma^{Born}_T=\frac{\lambda^4_D}{2\pi m^2_{\chi} v^4}\Bigg(ln(1+\frac{ m^2_{\chi} v^2}{m^2_\phi})-\frac{m^2_{\chi}v^2}{m^2_\phi+ m^2_{\chi}v^2}\Bigg)
	\end{equation} 	
	Outside the Born regime ($\lambda^2_D m_{\chi}/(4\pi m_\phi) \geq 1 $), there are two distinct regions {\it viz}, the classical regime and the resonance regime. In the classical regime ($\lambda^2_D m_{\chi}/(4\pi m_\phi \geq 1, m_{\chi} v/m_{\phi} \geq 1$), the solutions for an attractive potential is given by\cite{Tulin:2013teo,Tulin:2012wi,Khrapak:2003kjw}:	
	\vspace*{0.5cm}	
	\begin{equation}
		\sigma^{classical}_T =\left\{
		\begin{array}{l}			
			\frac{4\pi}{m^2_\phi}\beta^2 ln(1+\beta^{-1}) ~~~~~~~~~~~\beta \leqslant 10^{-1}\\
			\frac{8\pi}{m^2_\phi}\beta^2/(1+1.5\beta^{1.65}) ~~~~~~~~10^{-1}\leq \beta \leqslant 10^{3}\\
			\frac{\pi}{m^2_\phi}( ln\beta + 1 -\frac{1}{2}ln^{-1}\beta) ~~~~~\beta \geq 10^{3}\\
		\end{array}
		\right.
	\end{equation}  \\	
	where $\beta = 2 \lambda^2_D m_{\chi}/(4\pi m_\phi) v^2$	
	%Both the Born and the classical regime do not provide us the mild velocity dependence in the cross-section required to explain the small scale issues. However one interesting 	
	In the resonant regime ($\lambda^2_D m_{\chi}/(4\pi m_\phi) \geq 1, m_{\chi} v/m_{\phi}  \leq 1$), the quantum mechanical resonances and anti-resonance in $\sigma_T$ appear due to (quasi-)bound states formation in the attractive potential. In the resonant regime, an analytical formula for $\sigma_T$ is not available and one needs to solve the non-relativistic Schrodinger equation by partial wave analysis. Instead, here we use the non-perturbative results obtained by approximating the Yukawa potential to be a Hulthen potential $\Big( V(r) = \pm \frac{\lambda^2_D}{4\pi}\frac{ \delta e^{-\delta r}}{1-e^{-\delta r}}\Big)$, which is given by~\cite{Tulin:2013teo}:	
	\begin{equation}
		\sigma^{Hulthen}_T = \frac{16 \pi \sin^2\delta_0}{m^2_{\chi} v^2}
	\end{equation}
	where $l=0$ phase shift $\delta_0$ is given in terms of the $\Gamma$ functions by :
	\begin{eqnarray}
		\delta_0 &=arg \Bigg(i\Gamma \Big(\frac{i m_\chi v}{k~ m_\phi}\Big)\bigg/{\Gamma (\lambda_+)\Gamma (\lambda_-)}\Bigg)\nonumber\\
		\lambda_{\pm} &=
		\begin{array}{l}
			1+ \frac{i m_\chi v}{2 ~k ~m_\phi}  \pm \sqrt{\frac{\alpha_D m_\chi}{k m_\phi} - \frac{ m^2_\chi v^2}{4 k^2 m^2_\phi}}\\
		\end{array}
	\end{eqnarray}   
	and $k \approx 1.6$ is a dimensionless number.

	%\newpage
	
\providecommand{\href}[2]{#2}\begingroup\raggedright
\endgroup

%\bibliography{bibitems,refrhn,refn,ref,ref_old,ref1,ref2,ref3,refl,ref11,ref12,re13,ref14}
	%\bibliographystyle{apsrev}
	%\bibstyle{apsrev}
\bibliographystyle{JHEP}
%\bibliography{ref}
\end{document}